\documentclass[9pt]{elife}

\usepackage{amsmath}
\usepackage{booktabs}
\usepackage{textgreek}

\titleformat{\paragaph}{\slshape}{}{}{}

\newcommand{\phiA}{\phi_{A}}
\newcommand{\phiD}{\phi_{B}}

\newcommand{\fig}{figure}

\newcommand{\eqn}{equation}

\begin{document}
\title{Rheological basis of skeletal muscle work loops}
\author{Khoi D. Nguyen}
\author[1*]{Madhusudhan Venkadesan}
\affil{Department of Mechanical Engineering \& Materials Science, Yale University, New Haven, CT, USA}

\corr{m.venkadesan@yale.edu}{MV}



\maketitle

\begin{abstract}
    Skeletal muscle is subjected to simultaneous time-varying neural stimuli and length changes \emph{in vivo}.
    Work loops are experimental representations of these \emph{in vivo} conditions and exhibit force versus length responses that are not explainable using either soft matter rheology or the classical isometric and isotonic characterizations of muscle.
    These gaps in our understanding have often prompted the search for new muscle phenomena.
    However, we presently lack a framework to explain the mechanical origins of work loops that integrates multiple facets of current understanding of muscle, as a rheological material and also a stimulus-responsive actuator.
    Here we present a new hypothesis that work loops emerge by splicing together force versus length loops corresponding to different constant stimuli.
    Using published muscle datasets and a detailed sarcomere model, we find that the hypothesis accurately predicts work loops and helps understand them in terms of rheological behaviors measured at fixed-stimuli.
    Importantly, this framework identifies conditions under which a rheological understanding of muscle fails to explain the emergent work loops, and new muscle phenomena may be necessary to explain its \emph{in vivo} function.
\end{abstract}

\section*{Introduction}
Rheology, or how materials respond to being deformed, is a central consideration for living materials \citep{Gardel2008aa,Kollmannsberger2011aa,Kasza2007aa}.
Many biological tissues may be considered tunable because their functionality arises from a modulation of their rheological properties by an external control parameter or a stimulus \citep{Banerjee2020aa}.
One tunable material of considerable relevance to animals \citep{Lindstedt2016aa,Nishikawa2018aa}, and the object of much engineering mimicry \citep{Gennes1997aa,Madden2004aa,Hines2017aa}, is skeletal muscle.
A skeletal muscle's rheological behavior is actively regulated by the nervous system and is crucial for how animals control their body movements \citep{Dickinson2000cg,Nguyen2018aa}.

There are several broad groups of experimental characterizations of skeletal muscle rheology that each provide a different cross-sectional view of muscle phenomena.
Isometric and isotonic measurements capture the steady forces developed by muscle as a function of its length and shortening velocity \citep{James1996aa,Zajac1989aa}.
Transient forces in response to step length, step velocity, and twitch perturbations capture the initial fast response and history-dependent relaxation to steady-state \citep{Sandercock1997aa,Rassier2004nt,Lakie2019vf,Huxley1971aa}.
Oscillatory rheological experiments, or sinusoidal analysis, measure the frequency-dependent storage and loss moduli under fixed neural or electrical stimulation and length oscillations \citep{Machin1960aa,Kawai1980aa}.
And lastly, work loop analysis capture the mechanical actuation of muscles when they are simultaneously subjected to time-varying stimulation and length oscillations \citep{Josephson1985aa,Ahn2012aa}. 
Underlying all these broad groups are biophysical models of actomyosin crossbridges that form the internal motor machinery of muscle \citep{Huxley1957aa,Huxley1971aa, Nguyen2021ab}.
In this paper, we hypothesize and test a connection between two of the broad groups, namely oscillatory rheological experiments and work loops.
Such connections between different rheological characterizations can help illuminate new understandings of muscle phenomena and, in this case, provide a rheological basis for the shape of skeletal muscle work loops.

Work loop analysis is prevalently applied to study a skeletal muscle's forces in response to an externally imposed oscillatory length perturbation while it is actively regulated by neural or electrical stimuli \citep{Josephson1985aa,Ahn2012aa}.
The trajectory obtained by plotting the imposed oscillatory perturbation with the recorded force response forms the work loop, which is a graphical signature of the dynamics and work-producing capabilities of skeletal muscles \citep{Nishikawa2018aa,James1996aa}.
The shape of work loops is an important determinant of the biomechanical functions that the muscle provides \citep{Dickinson2000cg}.
It depends on the precise timing of the stimulus, the frequency of oscillation, the muscle's physiological properties, and other factors that are still vigorously debated \citep{Josephson1985aa,James1996aa,Dickinson2000cg,Sawicki2015aa}.
As a result, we currently lack a cohesive framework to understand and predict the emergence of complex work loop shapes.
For example, two cockroach leg extensor muscles that appear nearly identical under isometric or isotonic force characterizations generate loops of markedly different shapes, implying different functional consequences to the cockroach \citep{Ahn2002aa,Ahn2006aa}.

\begin{figure}[tb]
    \center{}
    \includegraphics[width=\columnwidth]{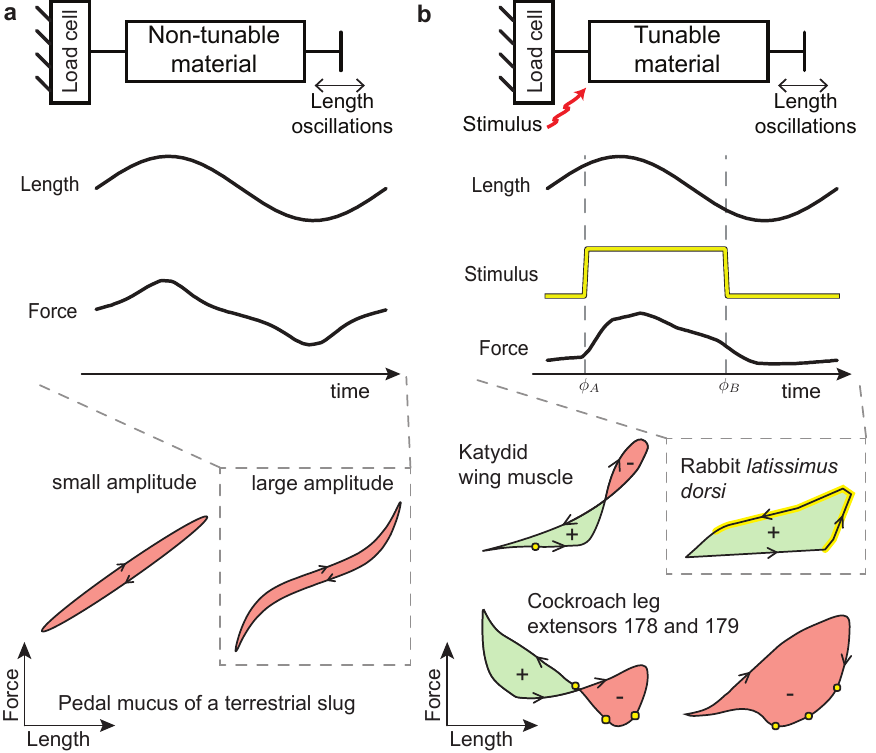}
    \caption{
    \textbf{Comparison of measured force-length loops in non-tunable and tunable materials.}
    \textbf{a}, Oscillatory rheology of a passive non-tunable material (pedal mucus of a terrestrial slug, {\it Limax maximus\/}) for small and large amplitudes \citep[adapted from][]{Ewoldt2008aa}. 
    \textbf{b},
    Work loops under time-varying stimuli of the wing muscle of a katydid ({\it Neoconocephalus triops\/}), rabbit {\it latissimus dorsi\/} muscle, and cockroach leg extensor muscles 178 and 179 \citep[adapted from][respectively]{Josephson1985aa,James1996aa,Ahn2002aa}.
    Yellow dots and thick yellow lines indicate discrete and continuous stimulation, respectively. The loops have been rescaled for visual comparison. Positive and negative mechanical work output are shaded green and red, respectively.
    All loops shown are taken from experimental measurements.
    }
    \label{fig:Intro}
\end{figure}

The difference between oscillatory rheological experiments and work loop analysis is that one involves fixed neural or electrical stimulation whereas the other involves time-varying stimuli.
A question that arises is if a detailed accounting of the time-varying stimuli permits a characterization of work loops as a superposition of rheological properties measured at fixed-stimuli.
The tunable nature and nonlinear rheological properties of skeletal muscles, however, are several hurdles.
For example, under fixed-stimuli, a muscle follows force-length trajectories that shares similarities with other soft and passive materials \citep[figure \ref{fig:Intro}a]{Tschoegl1989aa,Gennes1997aa,Madden2004aa} and can be non-elliptic in shape for large-amplitude which suggest a nonlinear oscillatory rheology.
And when the rheology is changing in time due to variable stimuli, the trajectory is far more complex and exhibits loop features like self-intersections and directional changes between clock-wise and counter-clockwise loops that passive materials do not show (figure \ref{fig:Intro}b).
Here, we build upon current fixed-stimuli rheological characterizations and admit tunability under a specific hypothesis and examine to what extent measurements from oscillatory rheological experiments can explain the emergent shape of skeletal muscle work loops.

In this paper, we develop and test the hypothesis that skeletal muscle work loops measured under time-varying neural or electrical stimulation emerge by transitioning between underlying fixed-stimulus force-length loops.
We call this transition between different loops as splicing, and term the framework developed here as the splicing hypothesis.
The paper starts with pedagogical preliminaries to introduce and define ideas of oscillatory rheology and linear viscoelasticity. 
We then present the mathematical formulation of the splicing hypothesis and derive a minimal parameterization for the space of all possible loop shapes obtained using a singular version of the splicing hypothesis where the transition between force-length loops is assumed to be instantaneous.
The space of loop shapes lends insights into different modes of mechanical actuation that can arise as a result of a skeletal muscle's tunable rheology and forms the premise for understanding the shapes that emerge under more gradual transitions.
We then show two parallel validations of our mathematical framework using published work loops of short-horn sculpin (\textit{Myoxocephalus scorpius}) abdominal muscles and using direct numerical simulations of a detailed biophysical sarcomere model.
We then discuss the implications of our framework for muscle modeling and experiments.

\section*{Preliminaries}
\subsection*{Oscillatory rheology}
Oscillatory rheology characterizes materials with invariant properties by generalizing upon notions of stiffness and damping to steady dynamical conditions \citep{Tschoegl1989aa,Hyun2011aa}.
To do so, the material's force (or stress) response to sinusoidal length (or strain) perturbations of different frequencies and amplitudes are characterized by force-length loops, otherwise known as Lissajous figures \citep{Lissajous1857aa}, that provide a graphical signature of the material's rheology \citep{Tschoegl1989aa}.
The force-length loops are approximately elliptic for small amplitude oscillations, but are typically non-elliptic for larger amplitudes (\fig~\ref{fig:Intro}a).
For small amplitudes, the complex modulus $E(\omega) = E'(\omega) + i\,E''(\omega)$ captures the material's dynamic response and generally depends on the oscillatory frequency $\omega$ \citep{Tschoegl1989aa}.

The storage modulus $E'(\omega)$ and loss modulus $E''(\omega)$ are respectively the in-phase and out-of-phase components of the measured force divided by the imposed length amplitude.
Numerous past experiments of muscle \citep{Kawai1980aa,Kawai1984yo,Campbell2010aa,Tanner2011aa, Palmer2020aa} and other natural and engineered tunable materials \citep{Gardel2008aa,Kollmannsberger2011aa, De-Vicente2011aa} characterize the measured force-length loops under a constant external stimulus using the storage and loss moduli.
Although {\it in vivo\/} muscle strain is often greater than the small amplitudes used in oscillatory tests, the loss and storage moduli have helped develop predictive models for muscle's dynamic response \citep{Rack1966aa,Palmer2007aa,Tanner2011aa} and guided the interpretation of {\it in vivo\/} data \citep{Niederer2019aa,Palmer2020aa}.

Non-elliptic force-length loops can occur in response to large amplitude oscillations that probe nonlinear rheological behaviors, for example, an inherent stress or strain dependence of the moduli, within muscles and other materials \citep[figure \ref{fig:Intro}a]{Tschoegl1989aa,Ewoldt2008aa,Hyun2011aa,Baxi2000aa}.
The nonlinearities are understood to be embedded within higher order terms of a Fourier expansion whose leading order terms are the storage modulus $E'$ and loss modulus $E''$\citep{Ewoldt2008aa,Hyun2011aa}.
We shall formulate our mathematical framework in terms of the leading order storage and loss moduli and focus on the linear oscillatory rheology of skeletal muscles because of the vast experimental literature that currently exist on them \citep{Kawai1980aa,Machin1960aa}.
The generalization to non-elliptic loops, however, is a direction extension of the framework by an inclusion of the higher order terms (see Appendix \ref{Appendix:Fourier}).

\textit{Sign convention:}
Following the muscle literature \citep{Josephson1985aa}, we take a sign convention in which increasing length is positive but positive forces imply the opposite sense, namely contraction.
So a positive or counter-clockwise loop is when the material performs work on the environment, and a negative or clockwise loop is when the material absorbs work (`$+$' and `$-$' regions in figures~\ref{fig:Intro} and~\ref{fig:Ellipse}). 

\section*{Mathematical framework}
\subsection*{Tunable oscillatory rheology}
Current rheological methods under a fixed stimulus do not directly accommodate the complex loop shapes that arise under a time-varying stimulus (\fig~\ref{fig:Intro}b for example).
We extend current methods and incorporate a time-varying element using a new hypothesis: splicing together fixed-stimulus force-length loops at different junctions predicts the shape of complex work loops. 

\begin{figure}[htb]
    \center
    \includegraphics[width=\columnwidth]{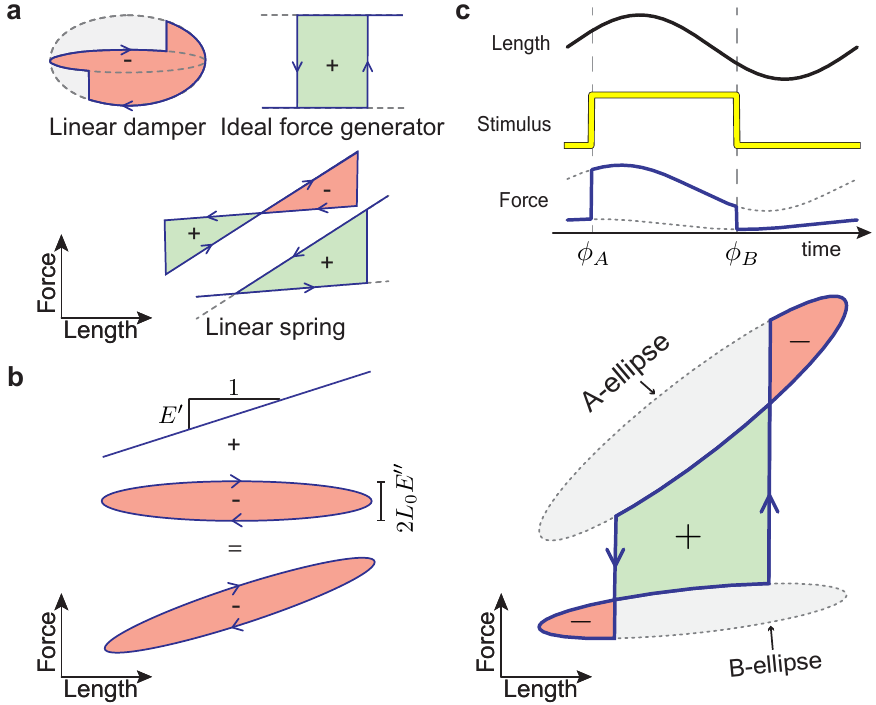}
    \caption{
    \textbf{Sketches to illustrate the splicing hypothesis.}
    \textbf{a,} Hypothetical work loops for idealized tunable elements.
    \textbf{b,} The force-length loop for a linear material is a sheared ellipse, which is the sum of a sloped line (elastic component) and a clockwise horizontal ellipse (viscous component).
    \textbf{c,} Splicing the state A ellipse and state B ellipse results in a new work loop.
    }\label{fig:Ellipse}
\end{figure}

We first illustrate the hypothesis by way of graphical examples and then state its mathematical formulation. 
Consider three idealized tunable elements, a Hookean spring with tunable stiffness and neutral length, a Newtonian damper with tunable viscosity, and an ideal force generator with tunable force output. 
The force generator accommodates the isometric forces a muscle exert when stimulated.
The force-length loops for these idealized elements under a constant stimulus are a sloped line for the Hookean spring, a horizontal clockwise ellipse for the Newtonian damper, and a flat line for the ideal force generator.
Periodically switching the properties of the idealized elements---stiffness, neutral length, damping, or force level---between two set of constant values or changing the timing of switching result in more complicated loops that exhibit reversals and self-intersections (\fig~\ref{fig:Ellipse}a).

We can generalize the idea from the simpler idealized elements to tunable linear viscoelastic materials by mapping stiffness to a storage modulus and viscosity to a loss modulus.
Combining the two moduli with the ideal force results in a vertically sheared ellipse (\fig~\ref{fig:Ellipse}b). 
The storage modulus controls the amount of vertical shear, the loss modulus controls the enclosed area, and the ideal force shifts the entire figure up or down.
Thus, under a time-varying stimulus that modulates all three rheological properties, a complex force-length loop emerges by switching between two sheared ellipses, each of which correspond to a distinct rheological state.
Consider a tunable material that has a greater storage modulus, loss modulus, and ideal contractile force at rheological state $A$ compared to state $B$, implying a more inclined, wider, and vertically offset ellipse for $A$ than $B$ (\fig~\ref{fig:Ellipse}c). 
Periodically switching the rheological states from $B$ to $A$ and back to $B$ would result in a splicing of the two ellipses to form the more complicated loop.
For this illustration, assume that the time taken to switch and settle into the new rheological state is negligible compared to the period of the oscillation.
Generically, there are two target points to jump to upon switching, but only one will traverse the loop in a direction consistent with its loss modulus, thus fully defining a new spliced loop built up from underlying sheared ellipses.
This spliced loop exhibits self-intersections and net positive work although both the A and B ellipses are individually dissipative.

\subsection*{Splicing hypothesis}
\label{sec:Splicing}
We state here the mathematical formulation of the hypothesis.
When phasically stimulated during $\phiA \le \omega t \le \phiD$, for oscillatory frequency $\omega$, time $t$, junctions $\phiA$ and $\phiD$ of switching between rheological states, the force response $F(t)$ to a sinusoidal length perturbation $L(t) = \Delta L \sin(\omega t)$ of amplitude $\Delta L$ is found by splicing the forces $F_B(t)$ and $F_A(t)$ corresponding to the two different rheological states $A$ and $B$, respectively, and expressed as,
\begin{equation}
    F(t) =  \begin{cases}
        F_{A}(t) & \text{for}\  \phiA \le \omega t \le \phiD,\\
        F_{B}(t) & \text{otherwise.}
    \end{cases}
    \label{eqn:splicedforces}
\end{equation}
The force-length loops obtained by graphing $F_A(t)$ and $F_B(t)$ against $L(t)$ introduce the notion of loops that are produced under a constant rheological states $A$ and $B$.
Given the loops corresponding to $F_A(t)$ and $F_B(t)$, the force response is completely described by a vector $[\phi_A,\phi_B]$ and the loops form a basis for all possible force responses that can arise from different choices of phasic timing.
We refer to these loops as basis loops.
In addition to the rheological state, these basis loops depend on the oscillation frequency and amplitude.
They are elliptic for linear materials but could generally be non-elliptic.
The spliced loop can be interpreted as an emergent periodic orbit of a piecewise smooth dynamical system (\fig~\ref{fig:Geometric}).
The tunable material's rheological states, the basis loops, are the constitutive pieces of the piecewise system.
We refer to the construction of work loops from basis loops as splicing (\fig~\ref{fig:Ellipse}c, \eqn~\ref{eqn:splicedforces}).

\subsection*{Shape-space of spliced loops}
\label{sec:nondimensional}

Shape is an important feature of work loops and directly embeds not only the net mechanical work a tunable material performs but also the type of mechanical actuation the material provides.
Specifically, in the case of muscle \citep{Dickinson2000cg}, whether it performs mechanical work as a tunable spring, tunable dashpot, or tunable ideal force generator has vastly different biomechanical implications even though the net mechanical work performed may be identical \citep{Nguyen2018aa}. 
A tool that the splicing approach affords is a parameterization of loop shape in terms of oscillatory rheological characterizations and therefore also a rheological basis for the mechanical actuation that muscle provides based on loop shape.
We show here the derivation of the parameterization and present a two-dimensional space of all possible loop shapes that can arise from splicing together basis loops.

For this derivation, we use linear rheological response of tunable materials where the material at a constant rheological state is fully described by an ideal force term, and the storage and loss moduli.
The analysis for nonlinear rheological responses is a direct extension of the linear theory but adds additional storage and loss moduli parameters that correspond to higher harmonics (Methods \ref{Appendix:Fourier}). 
The oscillatory force response $F(t)$ of a tunable material that is phasically stimulated between rheological states $A$ and $B$ depends on the ideal force terms $F_{A0}$ and $F_{B0}$, the storage moduli $E'_A$ and $E'_B$, the loss moduli $E''_A$ and $E''_B$, and length amplitude $\Delta L$, according to,
\begin{equation}
    F(t) = \begin{cases}
        F_{A0} + \Delta L\ \left(E_A' \sin\omega t + E_A'' \cos\omega t\right),
        \hfill\ \text{for}\ \phiA \le \omega t \le \phiD,\\
        F_{B0} + \Delta L\ \left(E_B' \sin\omega t + E_B'' \cos\omega t\right),
        \hfill\ \text{otherwise}.
    \end{cases}
    \label{eqn:dimensional}
\end{equation}

\begin{figure}[htb!]
    \center
    \includegraphics[width=\columnwidth]{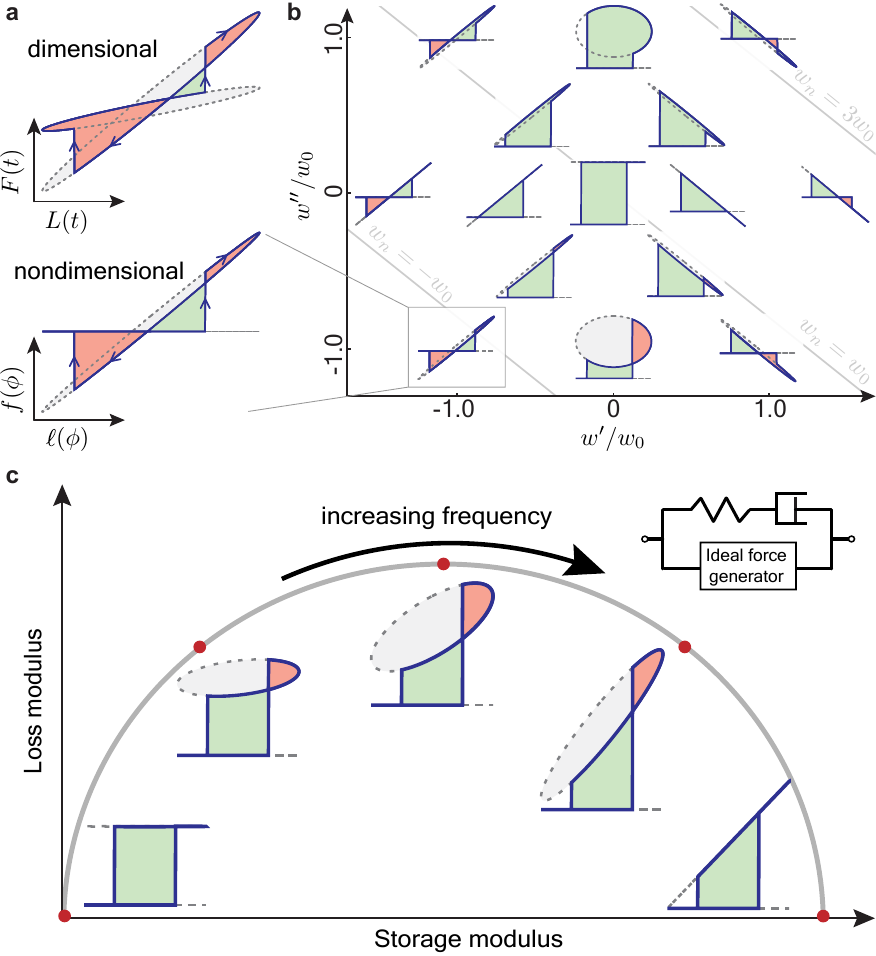}
    \caption{
    \textbf{Shape-space of spliced loops.}
    \textbf{a,} Representative loops to illustrate nondimensionalization of loop shape. 
    \textbf{b,} Two work ratios define the shape-space of spliced loops, one measures the net mechanical work by variable elastic modulus and the another the work by variable loss modulus. The loops are scaled to have the same width and height and shown here for one stimulation protocol ($\phi_A = \pi/6$ and $\phi_D = 5\pi/4$). Contour (gray) lines of net work $w_n$ are also shown.
    All loops shown are computed using \eqn~\eqref{eqn:nondimensional}.
    \textbf{c,} Example of traversing the shape space for an ideal force generator in parallel with a tunable Maxwell body, a spring and dashpot in series, which exhibits vastly different loop shapes simply by changing the oscillatory frequency (Methods \ref{Methods:Maxwell}).
    }
    \label{fig:LoopSpace}
\end{figure}

Atypical loop shapes, which are not seen in nontunable materials, emerge from phasic changes in rheological states.
Therefore, it is the difference between the two rheological states that introduces new loop features (\fig~\ref{fig:LoopSpace}a).
We subtract the force response $F_B(t)$ from $F(t)$, and derive nondimensional expressions using the length scale $\Delta L$, force scale $(F_{A0}-F_{B0})$, and timescale $1/\omega$.
In terms of the nondimensional phase $\phi=\omega t$, the stimulated nondimensional force response $f_A(\phi) = (F_A(t)-F_B(t))/(F_{A0}-F_{B0})$ is expressed using the difference in moduli $\Delta e' = \Delta L (E_A'-E_B')/(F_{A0}-F_{B0})$ and $\Delta e'' = \Delta L (E_A''-E_B'')/(F_{A0}-F_{B0})$ as,
\begin{equation}
    f_A(\phi) = 1 + \Delta e' \sin\phi + \Delta e'' \cos\phi.
    \label{eqn:nondimensional fA}
\end{equation}

We recast the modulus parameters $\Delta e'$ and $\Delta e''$ in terms of the nondimensional work $w'$ and $w''$ that is performed by the material as a result of switching the storage and loss moduli respectively, and an additional work $w_0$ as a result of switching the ideal force component between $F_{B0}$ and $F_{A0}$.
Thus, the nondimensional sinusoidal length perturbation is $\ell(\phi) = \sin(\phi)$ and the force response $f(\phi)$ is
    \begin{align}
        \nonumber f(\phi) &= \frac{F(t) - F_B(t)}{F_{A0}-F_{B0}}\\
        &=
        \begin{cases}
            1   + \frac{w'}{w_0}\, \frac{2}{\ell_A+\ell_B}\sin\phi + \frac{w''}{w_0}\, \frac{\ell_B-\ell_A}{\beta} \cos\phi,\\
            \hfill \text{for}\ \phiA \le \phi \le \phiD,\\
            0,\ \text{otherwise}.
        \end{cases}
        \label{eqn:nondimensional}
    \end{align}
The shape factor $\beta=\int_{\phiA}^{\phiD}\cos^2\phi\, d\phi$ represents the partial area of the loss-ellipse traversed in switching the rheology, and $\ell_A = \ell(\phiA)$ and $\ell_B = \ell(\phiD)$ are the dimensionless lengths at switching from $A$ to $B$ and then from $B$ back to $A$, respectively.
The nondimensional work $w'$ for switching the elastic modulus, $w''$ for switching the loss modulus, and $w_0$ for switching the ideal force term are,
    \begin{align}
        w' &= -\Delta e' \int_{\ell_A}^{\ell_B} \ell d\ell = \frac{1}{2}\Delta e'(\ell_A^2 - \ell_B^2),
        \label{eqn:Wsprings}
        \\
        w'' &= -\Delta e''\int_{\phiA}^{\phiD} \cos^2\phi d\phi = -\Delta e'' \beta 
        \label{eqn:Wloss}
        \\
        w_0  &=  -\int_{\ell_A}^{\ell_B} d\ell = \ell_A - \ell_B,
        \label{eqn:Wideal}
    \end{align}
and the nondimensional and dimensional net work per loop are respectively,
    \begin{align}
        w_{n} &= \oint fd\ell  =   w_0+ w' + w'',\ \text{and}
        \label{eqn:Wnet}
        \\
        W_n &= w_n \Delta L (F_{A0}-F_{B0}) - \pi E''_B \Delta L^2.
        \label{eqn:dimensional net work}
    \end{align}

Expressing the force response, and thus the loop shape, using work ratios aids in the interpretation of the response under time-varying stimuli in terms of the material's functionality as a force actuator, elastic body, or viscous damper (\fig~\ref{fig:LoopSpace}b).
The choice of shape parameterization also delineates the rheological states $A$ and $B$ from the effect of the stimulus protocol that are captured by the parameters $(\ell_B+\ell_A)/2$,\ and $\beta/(\ell_B-\ell_A)$.
Current approaches to functionally interpreting skeletal muscle work loops are based on qualitative observations about the positive and negative regions of the loops \citep{Josephson1985aa,Ahn2012aa}.
The shape-space provides a quantitative means to understand the net effect of the shape in terms of relative work contributions from the storage and loss moduli underlying the rheological response relative to the ideal force term.
Therefore, the coordinates of a loop in the shape-space provide a quantitative functional interpretation of the loop.

Although the veracity of the shape-space remains to be tested, the splicing approach provides a single framework in which to view vastly different loop shapes.
For example, different loop shapes may arise from the same material simply by varying the oscillatory frequency and holding all else constant (\fig~\ref{fig:LoopSpace}c).
This is because the material's rheological properties are frequency-dependent.
The splicing approach enables the application of rheological modeling to generate predictions for skeletal muscles when it is subjected to varying stimuli.
It may also prove to be a tool for designing actuators and programmable mechanical interfaces from tunable materials other than skeletal muscles because it is now possible to predict the range of loop shapes a material can exhibit and consequently the exchange of mechanical work with its environment. 

\section{Results}

The splicing hypothesis and its constitutive pieces are tested here using published muscle data and numerical simulations. We first show direct evidence for the passive basis loops in rat papillary muscles by graphically overlaying their force-length loops when deactivated and when subjected to a single electrical spike. We then test whether the splicing hypothesis accurately predicts loop shape using a dataset of work loops collected from short-horn sculpin (\textit{Myoxocephalus scorpius}) abdominal muscle. These work loops differ only in the phasic timing of an electrical stimulus and, according to the splicing hypothesis, arise from the same set of basis loops. Because the rat and sculpin work loops were not specifically collected to test splicing, we also implement a biophysical model of the sarcomere that incorporates calcium activation to compare predictions of the splicing hypothesis with computed work loops. 

In connecting theory with data and simulations, it is important to define stimulus and rheological states in the context of muscles.
The stimulus is an external control parameter that varies muscle behavior and can take different forms depending on experimental protocol.
It may be calcium concentration, binding affinity of actomyosin crossbridges, frequency of neural inputs, or frequency of electrical spikes that directly modulate the motor machinery of muscle.
The dynamics of engagement of the motor machinery, specifically the bond distributions of actomyosin crossbridges \citep{Nguyen2021ab}, is the muscle's rheological state that underlies its viscoelastic resistance to length perturbations.
The distinction between stimulus and rheological state introduces several complications in examining muscle data.
An example to be encountered with sculpin work loop data is that the transition between rheological states is slower than changes in the stimulus.
The stimulus may also be a function of length perturbations because of stretch-induced calcium release or heat production and the notion of a fixed-stimuli rheological state during measurement becomes less well-defined \citep{Sandercock1997aa,Hill1938aa}.
Furthermore, because there are multiple possible stimuli, holding one constant does not imply that the others are constant. So a fixed-stimulus rheological state is not one of all possible stimuli being constant, but one in which the stimulus under experimental control is held constant.
Recognizing all these complexities about muscle, the question we ask in examining work loop data and sarcomere models is if a superposition of rheologies measured at fixed-stimuli adequately captures the tunable rheology measured for time-varying stimuli.

\subsection*{Rat papillary work loops}
Work loops of rat papillary muscles show evidence for one of the basis loops corresponding to a low stimulus, namely the analog to the B-ellipse in \fig~\ref{fig:Ellipse}c that has a weaker rheological response compared to the A-ellipse.
Although papillary muscles are cardiac rather than skeletal muscles, they are similar based on the fact that they both are underlay by the mechanics of sarcomeres.
Baxi et al.\ \citep{Baxi2000aa} overlays work loops measured for a single electrical impulse with passive force-length loops of rat papillary muscles on the same plot.
The work loop closely traces the passive loop up to onset of the electrical impulse (\fig~\ref{fig:FishWorkLoops}a), providing direct evidence which shows that the passive loop is a basis loop corresponding to an unstimulated papillary muscle.
But Baxi et al.\ \citep{Baxi2000aa} did not measure a basis loop when the papillary muscle was stimulated.
This could be because stretching highly activated muscles often induces tissue damage \citep{Lindstedt2001aa}, thus making it experimentally infeasible to directly measure basis loops of an activated muscle at large strains.

\subsection*{Sculpin work loops}
\begin{figure}[htbp!]
    \center
    \includegraphics[width=\columnwidth]{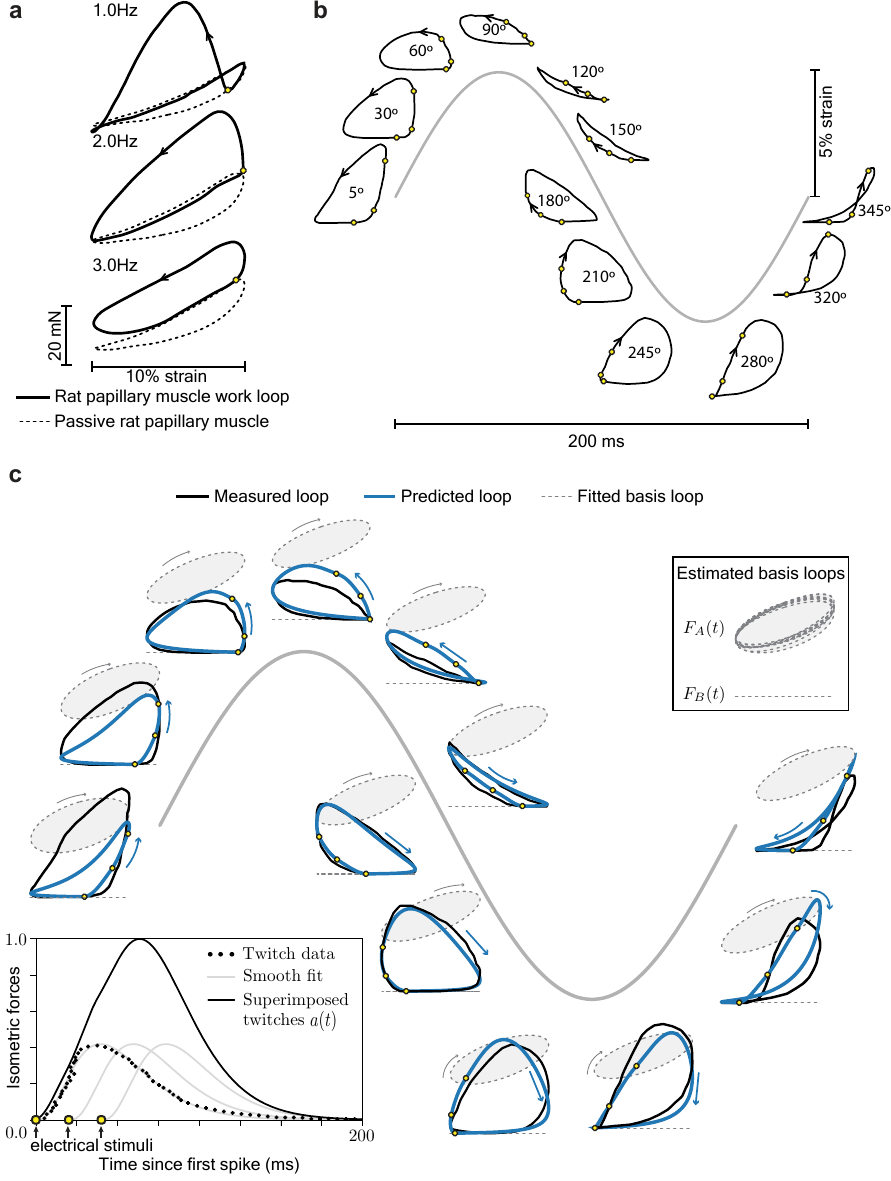}
    \caption{
    \textbf{Work loops, basis loops, and gradual transitions.}
    \textbf{a,} Passive force-length loops and works loops of rat papillary muscle show that measured work loops trace closely along the passive loop up to the electrical spike (yellow dot) that actives the muscle's motor machinery \citep[adapted from figure 3 of][]{Baxi2000aa}.
    \textbf{b,} Work loops from a short-horned sculpin abdominal muscle measured at 5Hz and 5\% strain amplitude and for different stimulation protocols \citep[adapted from figure 4 of][]{Johnson1991aa}.
    Three electrical spikes were applied in each cycle (yellow dots).
    Work loop locations on the sine wave indicate the phase difference between the first stimulus and the length oscillation. 
    \textbf{c,} A leave-one-out analysis to estimate the stimulated basis loop (grey shaded ellipses), used to predict the work loop (blue), and compare with measurement (black).
    The slow transition in rheological states is modeled using the isometric twitch response to a single electrical spike to construct an interpolation variable $a(t)$. The twitch response was found from separate measurements of the short-horned sculpin abdominal muscle (bottom-left inset) \citep[adapted from figure 1b of][]{Altringham1988aa}.
    }\label{fig:FishWorkLoops}
\end{figure}

Although direct measurements of basis loops under high neural or electrical stimulus are not currently available, we show a means to infer them by pooling data from multiple work loop measurements of the same muscle.
By using the inferred basis loops, we show how to apply splicing and predict work loops with which to compare with measured work loops.
For this analysis, we use a previously published dataset consisting of twelve work loops from a short-horn sculpin (\textit{Myoxocephalus scorpius}) muscle \citep{Johnson1991aa} subjected to a 5\% strain amplitude and a 5Hz oscillatory frequency.
In that study, three successive electrical spikes spaced 20\,ms apart were applied per cycle and the onset of the first stimulus relative to the length oscillation was systematically varied from 0$^\circ$ to 360$^\circ$ phase lag to produce twelve work loops that vary considerably in their shape (\fig~\ref{fig:FishWorkLoops}b). 
Because the twelve loops differ only in stimulus timing, the sculpin dataset provide a testing ground of the splicing hypothesis because they are, in theory, underlay by the same set of basis loops.

A complication arises that unlike the simplified illustration of the splicing approach (\fig~\ref{fig:Ellipse}), the effect of the stimulus on muscle is not instantaneous nor is it constant for the entire duration.
Instead, work loop measurements typically apply impulsive neural or electrical spikes that cause a gradual rise and fall in intracellular calcium concentration, which in turn leads to a gradual engagement and disengagement of the muscle's motor machinery.
The degree of engagement of the motor machinery is the rheological state that affects force production, but it is often experimentally inaccessible or difficult to quantify.
We overcome this inaccessibility and approximate the rheological state using independent experimental measurements of the force twitch response to a single electrical spike under isometric conditions \citep{Altringham1988aa}.
Because the length is constant, the time-course of the force development and decay reflects the engagement of the motor machinery.
The pulse duration in the isometric twitch study was 1\,ms, different from 2\,ms that was used in the sculpin dataset.
But both are much smaller than the 20\,ms inter-spike interval, the $\sim$100\,ms force rise and relaxation time, or the 200\,ms time period of the oscillation.
So we treat the stimuli in both studies as impulses.
Furthermore, the work loop protocol used three successive electrical spikes.
We find the total response to three spikes using a superposition of three twitch responses and rescale the response to lie between zero and one to yield a normalized interpolation variable $a(t)$ (bottom left inset of \fig~\ref{fig:FishWorkLoops}c, and Methods \ref{Methods:Sculpin}) that embeds within it the dynamics between stimuli and rheological states.
This procedure uses the single-spike response to construct the response to more generalized inputs, three spikes in this case, and is similar to past applications of impulse responses in neural systems involving sensory \citep{Gupta2015aa} and motor activation dynamics \citep{Weiss1988aa}.

We modify the splicing approach of \eqn~\eqref{eqn:splicedforces} to accommodate the sculpin muscle's continuously varying rheological state by using the normalized interpolation variable $a(t)$. 
The geometric picture for this modification is that the measured force reflects an intermediate rheological state that a muscle transiently passes through at any point in time.
The interpolation variable $a(t)$ specifies this intermediate rheological state such that when $a(t)$ gradually changes from zero to one, the sculpin work loop transitions from one basis loop corresponding to $F_B(t)$ to another corresponding to $F_A(t)$.
Using standard practice to linearly relate forces \citep{Zajac1989aa}, and for an onset time $t_0$ when the stimulus is first applied, the mathematical representation of this modification of \eqn~\eqref{eqn:splicedforces} to permit a slow transitioning between rheological states $A$ and $B$ is
\begin{align}
    F(t) = a(t-t_0)F_A(t) + (1-a(t-t_0))F_B(t).
    \label{eqn:force interpolation}
\end{align}
The measured passive loop is generally non-zero much like the Baxi et al.\ study \citep{Baxi2000aa}, but we assume it to be zero in the sculpin dataset because the stimulated loop $F_A(t)$ appears to be the dominant rheological structure.
Furthermore, we parameterize the stimulated basis loop using an ideal force term $F_{A0}$, storage modulus $E'_A$, and loss modulus $E''_A$.
Thus, for an applied length oscillation of amplitude $\Delta L$, frequency $\omega$ and a stimulus onset time $t_0$, the predicted spliced response is simplifies from \eqn~\eqref{eqn:force interpolation} by taking $F_B(t) = 0$ as
    \begin{align}
        F(t) = a(t-t_0)(F_{A0} + \Delta L E_A' \sin\omega t + \Delta L  E_A''\cos\omega t). \label{eqn:force interpolation fish}
    \end{align}
The interpolation variable $a$, stimulus onset time $t_0$, and amplitude $\Delta$ are all externally specified which leaves the three rheological parameters ($F_{A0}$, $E'_A$, and $E''_A$) of the stimulated basis loop to be determined.
In the more general case where the passive loop cannot be ignored, the passive loop adds three additional rheological parameters to be determined.

We estimate the three parameters ($F_{A0}$, $E'_A$, and $E''_A$) by fitting the sculpin work loop data to \eqn~\eqref{eqn:force interpolation fish} in a leave-one-out analysis to test the splicing hypothesis (Methods \ref{Methods:Sculpin}).
Briefly, the analysis excludes one work loop from the dataset of twelve loops and uses the remaining eleven to estimate the three parameters.
The estimated parameters are then used to generate a prediction that is compared that with the excluded loop (\fig~\ref{fig:FishWorkLoops}c).
We repeat this procedure for all twelve loops, generating a different triplet ($F_{A0}$, $E'_A$, and $E''_A$) for each and avoiding a tautological use of the dataset in which a measured loop is used to generate its own prediction.

We quantitatively compare the loops generated by the leave-one-out analysis and the measured sculpin work loops using Pearson's correlation coefficient and net mechanical work performed as goodness of fit measures.
These measures are plotted in \fig~\ref{fig:FishWorkLoopComparison} as functions of the stimulus timing.   
The Pearson's correlation coefficient compares the predicted force response and the measured force response (see Methods \ref{Methods:Sculpin}), and a value close to one implies a strong correlation. We find that it ranges from 0.90-0.99 for all loops. 
The net mechanical work performed is a standard measure in work loop analysis to compare loops \citep{James1996aa,Altringham1990aa,Askew1997aa}, and we find it to be similar for all pairs of measured and predicted loops except for the 5$^\circ$ and 30$^\circ$ loops.
The estimated stimulated basis loop, \textit{i.e.} the fitted triplet ($F_{A0}$, $E'_A$, and $E''_A$), can also serve as a quantitative comparison and we find it to be almost unchanging across all twelve instances of the leave-one-out fitting procedure (top-right inset of \fig~\ref{fig:FishWorkLoops}c, \ref{tab:fitting ellipses to work loops}).
Based on these comparisons, we find that the predicted loops generated by the splicing hypothesis accurately captures the experimentally measured sculpin work loops.

Future investigation is needed, but we speculate that the slightly worse predictions when the stimulation phase is near $0^\circ$ might be due to stretch-induced doublet-potentiation in muscle where the effects of a stimulus are modified \citep{Sandercock1997aa}.
Briefly, stretch-induced doublet potentiation is the experimental observation that the muscle's force response to stretch sum up more than linearly when subjected to multiple back-to-back electrical spikes \citep{Nishikawa2018aa,Sandercock1997aa}. By comparison, our construction of the interpolation variable $a(t)$ assume that the force responses adds up linearly and, in doing so, our approach implicitly serves as a control to test nonlinear summations of forces.
Thus, in addition to revealing the rheological origins of muscle work loops, the splicing hypothesis may help identify circumstances when tissue-specific phenomena such as doublet-potentiation become functionally consequential.

\subsection*{Muscle biophysical model}
Although the rat and sculpin datasets provided two independent lines of experimental validation based on measured work loops, the datasets were not specifically collected to test splicing and represent only two muscles out of a myriad of muscle types. So to further test splicing, we use direct numerical simulations of a detailed biophysical model of the contractile machinery in muscle \citep{Walcott2014aa}. The objectives are two-fold. First, to present a work flow that guides design of future experiments in inferring the basis loops and interpreting measured work loops in terms of them. Second, to show that the splicing approach can predict loop shape in current muscle models and therefore show the applicability of splicing across multiple muscle types and not just of the muscles considered here. 
In light of these objectives, the biophysical model was chosen based on specific features that test the assumptions underlying the splicing hypothesis.
The biophysical model exhibits non-elliptical basis loops and has slow transitional dynamics between rheological states, and the precise value and interpretation of the model parameters are as suggested in the literature.

\begin{figure}[htbp!]
    \center
    \includegraphics[width=\columnwidth]{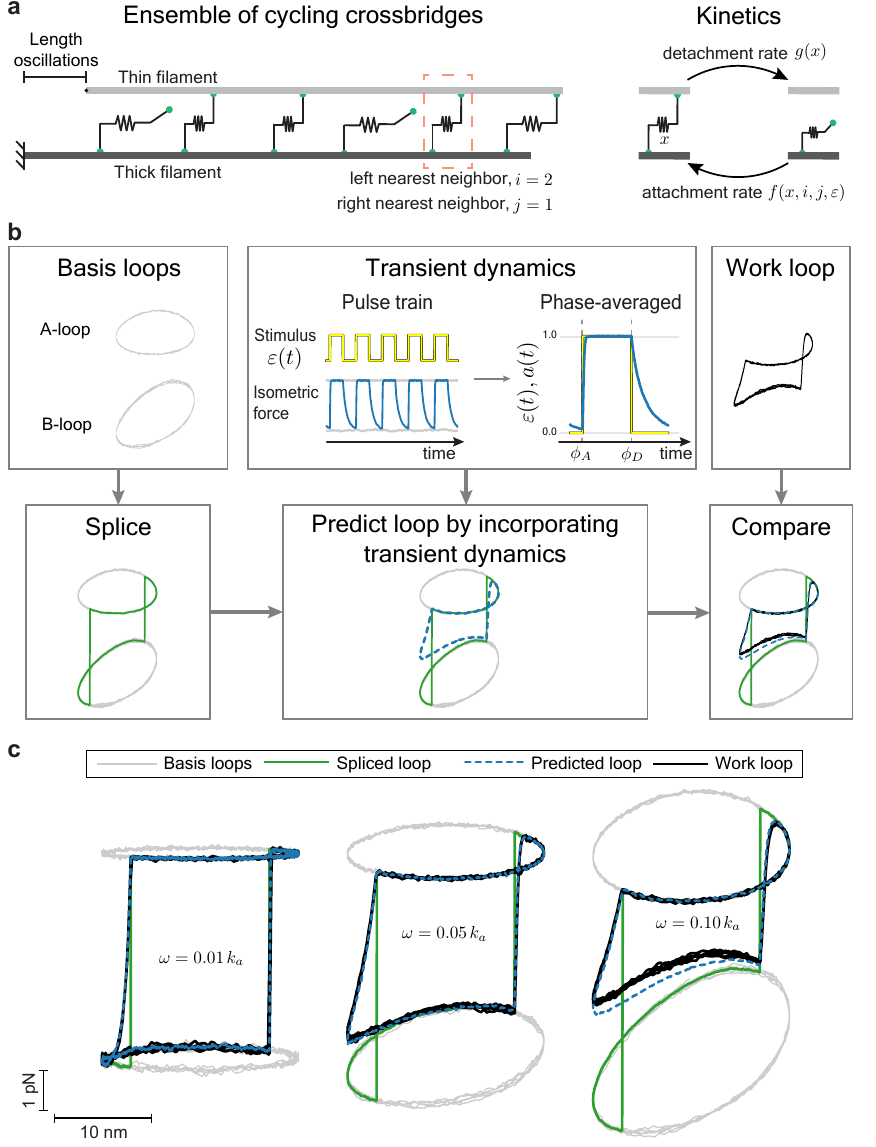}
    \caption{
    \textbf{Work loop prediction in muscle: a model-based validation of splicing.}
    \textbf{a,} The sarcomere is modeled as an ensemble of crossbridges that cycle between attached and detached states depending on its internal displacement $x$, distance to its left nearest neighbor $i$, distance to its right nearest neighbor $j$, and a coupling strength $\varepsilon$ that is varied as the external control parameter and used as the stimulus \citep[][Methods \ref{Methods:MonteCarlo}]{Walcott2014aa}. Loops are generated using Monte Carlo simulations of an ensemble consisting of 500,000 crossbridges.
    \textbf{b,} Flow chart to predict work loops based on basis loops and isometric responses, and thus assess the splicing hypothesis.
    \textbf{c,}
    Outputs of the flow chart for three different frequencies placed next to each other for comparison. 
    The basis loops and work loops are the direct results of Monte Carlo simulations whereas the spliced and predicted loop are computed using the splicing hypothesis.
    The frequency scale $k_a$ is the myosin attachment rate taken to be around $40Hz$ in chicken pectoralis at $25^\circ $C \citep{Walcott2014aa}.
    The vertical scale is in units of tension per crossbridge whereas the horizontal scale is in units of sarcomere displacement.
    }
    \label{fig:MonteCarlo}
\end{figure}

The detailed biophysical model by Walcott captures the forces of single sarcomeres, contractile structures of approximately  2.5\textmu m in length and which repeat in series to form a single muscle fibril.
Skeletal muscles comprise of bundles of such fibrils.
Within a single sarcomere, contraction is powered by myosin motors that interact with actin filaments by forming force-bearing crossbridges when myosin stochastically binds to actin, strokes and unbinds (\fig~\ref{fig:MonteCarlo}a).
Walcott's model incorporates realistic aspects such as thin filament activation dynamics and spatial coupling between actomyosin crossbridges \citep{Farman2010aa,Reconditi2017aa} in addition to the latest advances in modelling the Lymn-Taylor actomyosin crossbridge cycle that underlies force production in muscle \citep{Nishikawa2018aa,Lymn1971aa}.

We assess the splicing hypothesis in the sarcomere model by predicting work loops from basis loops using \eqn~\eqref{eqn:force interpolation}. The work loop $F(t)$ is obtained from the sarcomere model for a periodic stimulation protocol that varies between two values and the basis loops, $F_A(t)$ and $F_B(t)$, are obtained by holding the stimulus constant at each value. 
The stimulus used as the external control parameter is a coupling strength $\varepsilon$ that modulates the spatial coupling between neighboring crossbridges [Methods \ref{Methods:MonteCarlo}, \fig~\ref{fig:MonteCarlo}a], and the rheological state of the sarcomere model emerges from the crossbridges' response that make up the sarcomere's viscoelastic resistance to mechanical perturbations. 
We implement a periodic back and forth switch, \textit{i.e.} a rectangular pulse train, for the stimulation protocol in which the coupling strength $\varepsilon$ switches between values corresponding to pCa $= 6.17$ and pCa$=7.04$ [See Methods \ref{Methods:MonteCarlo}]. 
Such stimulation protocols are common when applying tetanic electrical stimuli \citep[e.g.][]{James1996aa}.
Much like the rat and sculpin datasets, the rheological state does not change instantaneously in response to the stimulus and has additional dynamics that must be accounted for.
Unlike for a model, the rheological state is generally not accessible in work loop experiments.
So, like for the sculpin dataset, we use the isometric response to the stimulation protocol and construct a linear interpolation variable $a(t)$ that parameterizes the rheological state (see \eqn~\ref{eqn:force interpolation}).
Specifically, the isometric response is collected over several periods of the stimulation protocol and averaged at the same angular phase of a period from 0$^\circ$ to 360$^\circ$., \text{i.e.} phased-averaged, and then normalized between zero and one to form the interpolation variable $a(t)$ at every time point in the period.
This approach uses only experimentally accessible elements and therefore the flow chart of \fig~\ref{fig:MonteCarlo}b can be applied to muscle work loop experiments.
The elements are: the basis loops obtained by holding the stimulus constant, the isometric force response to a time-varying stimulation protocol, and work loops obtained for the same stimulation protocol.

We also probe the model's nonlinearities to demonstrate that splicing can work even for non-elliptic basis loops by using peak-to-peak oscillatory amplitudes that are twice the power stroke length of myosin.
We repeat this for three different oscillation frequencies to elucidate the effect of slow versus fast transitions between rheological states relative to the oscillation time period  (\fig~\ref{fig:MonteCarlo}c).
To separately identify the roles of basis loops and dynamics associated with switching between them, we show both the spliced loop and final predicted loop.
The spliced loop assumes an instantaneous change in rheological states and the final predicted loop applies the interpolation variable estimated from isometric experiments (see \eqn~\ref{eqn:force interpolation} and panel titled transient dynamics in \fig~\ref{fig:MonteCarlo}b).
The predicted loops accurately reconstruct the work loops at all oscillatory frequencies and the idealized spliced loop is shown to underlie the overall shape of the work loops.
Not surprisingly, the sharp corners in the idealized splicing construction are smoothed out by incorporating the isometric force response which captures asymmetries in the timescales of force rise versus decay. 
If in muscle experiments, direct measurement of the basis loop could damage the tissue, we find that small amplitude measurements of basis loop could be used in lieu of the true large amplitude basis loop with little loss of accuracy, at least for this detailed biophysical model of muscle (\fig~\ref{fig:SmallAmplitude}).
Thus, work loops are comprised of two components: ({\bf i}) an idealized spliced loop formed from basis loops that can be found from the rheological response at constant stimulus, and ({\bf ii}) dynamics between transitioning between rheological states that smooth or regularize the idealized loop and can be found using isometric force responses.

\section*{Discussion}
In examining the mathematical framework and evidence presented here, the reader is alerted to some cautionary points.
Muscle is not monolithic and considerable physiological differences arise between different types of muscle; skeletal, cardiac, smooth, fast or slow, and many other varieties.
This paper tests the splicing hypothesis in two specific muscles from rat and a short-horn sculpin for which the type of data needed are presently available, but future studies may expand that set.
Furthermore, rheology is fundamentally a bulk property and only lends partial insight into the molecular mechanisms underlying it.
So rheological studies are complementary to ongoing studies and debates that are centered around the molecular mechanisms behind intriguing muscle phenomena such as history-dependence \citep{Sandercock1997aa,Rassier2004nt,Lakie2019vf}, length-dependent transitions \citep{Getz1998ny}, and other transient non-steady phenomena \citep{Herzog2000aa,Nishikawa2018aa}.
But, insofar as stable work loops can be measured and are applicable to mechanical actuations in animals, testing the splicing hypothesis will lend insight into the applicability of fixed-stimulus rheology to the functionally more realistic case of a time-varying stimulus.
In this manner, the work presented here takes a bottom-up and data-driven approach to assess how well a superposition of fixed-stimulus rheology explains the data under time-varying stimulation.
A thorough investigation into the mathematics of splicing together fixed-stimulus rheologies is related to the theory of piecewise dynamic systems, which could be an exciting direction to study muscle and potentially other active materials that violate assumptions underlying the modeling of passive rheological materials. 
Furthermore, the splicing hypothesis presents an analytical tool to incorporate steady-state rheology into predictions of non-steady conditions so that future investigations can unambiguously account for the role of steady-state rheology before attributing measured responses to new phenomena.

The quality of the loop predictions in the sculpin dataset and in the biophysical model point to the veracity of the splicing hypothesis and the rheological basis for the emergence of work loops in muscle.
They also underscores the robustness of two linearity assumptions, namely,
({\bf i}) linear rheology for the basis loop (ellipse-shaped), and 
({\bf ii}) the principle of superposition to extract an interpolation variable between rheological states from the isometric force responses, either single twitch responses or phase-averaged responses to a periodic stimulation protocol.
The robustness of the linear rheological assumption indicates that although muscle is a nonlinear material, ellipses or equivalently the frequency-dependent dynamic moduli can adequately capture the dominant features of the basis loops as exemplified by \fig~\ref{fig:SmallAmplitude} in which the small-amplitude response is used in lieu of the large-amplitude response.
So when length oscillations are non-sinusoidal, the basis loops can be expanded as a linear summation of dynamic moduli at multiple frequencies.
The robustness of the second linearity assumption, \textit{i.e.} \eqn~\eqref{eqn:force interpolation}, for constructing the transition between rheological states and using that to infer the behavior of a sub-maximally activated muscle is also to be expected based on proven antecedents in capturing neural stimulation dynamics in sensory systems \citep{Gupta2015aa} and studies on dynamic stiffness of ankle muscles \citep{Weiss1988aa}.
Thus we conclude that the basis loops are the building blocks for the work loop in muscle, in agreement with the splicing hypothesis.

While our focus is a rheological basis for skeletal muscle work loops, we speculate that the splicing hypothesis will be applicable to other striated muscles based on our examination of a detailed sarcomere model and the fact that an orderly arrangement of sarcomeres is found in all striated muscles.
Cardiac muscles are a key area of clinical relevance because their mechanical behavior is a central component of heart function and dysfunction \citep{Niederer2019aa,Fung1971aa}.
Further validation in more specialized cardiac muscle models and in experiments are certainly necessary and may be guided by the flow chart of \fig~\ref{fig:MonteCarlo}b. 
The basis loops are likely non-elliptic, as evidenced by rat papillary dataset \citep{Baxi2000aa} shown in \fig~\ref{fig:FishWorkLoops}a, and can be accounted for by introducing higher harmonic modes into the data analysis (see Methods \ref{Appendix:Fourier}).
The basis loops themselves may also be difficult to directly measure without inducing tissue damage, but can be inferred in an approach similar to the sculpin dataset analysis by pooling together multiple work loops that differ only in stimuli timing (see Methods \ref{Methods:Sculpin}) or using small-amplitude length oscillations as we did with the biophysical model (\fig~\ref{fig:SmallAmplitude}).
Keeping all these possible complexities in mind, the splicing hypothesis may provide a new phenomenological approach to modelling the rheological behavior of cardiac muscle in a manner that accounts for its tunability.

\section*{Conclusion}
We adapted current oscillatory rheological methods to admit tunable properties by splicing  fixed-stimulus rheologies and showed its predictive ability using published data on sculpin skeletal muscle and direct numerical simulations of a detailed sarcomere model.
We found that most but not all of the sculpin muscle's work loop is accounted for by an interpolation between basis loops and rheological states.
Our method incorporates experimental rheological characterization of muscle into predictions under time-varying neural or electrical stimuli, so that future investigations into new emergent muscle phenomena can account for and filter the effects of steady-state rheology.
In this manner, the splicing hypothesis is a new tool to study muscle mechanics and complements ongoing investigations into the molecular mechanisms that underlie the emergent rheological properties.
Finally, the shape-space of work loops provides a unified view of the vastly different loop shapes a muscle can exhibit to variably perform as motors, springs, dampers, and combinations thereof based on its tunable rheology.

\noindent\textbf{Acknowledgments:}
Funding support from the Raymond and Beverly Sackler Institute for Biological, Physical and Engineering Sciences at Yale, a National Institutes of Health training grant T32EB019941, and the Robert E. Apfel Fellowship awarded by Yale. 
This material is based upon work supported by the National Science Foundation under Grant No.\ 1830870.
M.M. Bandi, S. Mandre, and M. Murrell for helpful comments.

\noindent\textbf{Author Contributions:}
K.D.N. and M.V. conceived the research, designed the research, analyzed and interpreted the results, and wrote the paper.

\noindent\textbf{Competing interests:}
The authors declare no competing interests.

\bibliography{references}

\newpage

\setcounter{secnumdepth}{2}
\renewcommand{\thesubsection}{\Alph{subsection}}
\titleformat{\subsection}
  {\large\bfseries}
  {\thesubsection}{0pt}{. #1}[] 

\section*{Methods}
\label{Methods}

\subsection{Expansion with higher harmonics:}
\label{Appendix:Fourier}
The generalization of splicing the fixed-stimulus basis loops to include higher harmonics is a straightforward extension of splicing ellipses by using a Fourier series for the force response.
The force response to sinusoidal length perturbations as given by \eqn~\eqref{eqn:dimensional} generalizes to
\begin{equation}
F(t) = 
    \begin{cases}
        F_{A0} + \Delta L \sum\limits_k (E_{A,k}' \sin(k \omega t) + E_{A,k}'' \cos(k \omega t)), &  \text{for } \omega t \in [\phiA, \phiD]\\
        F_{B0} + \Delta L \sum\limits_k (E_{B,k}' \sin(k \omega t) + E_{B,k}'' \cos(k \omega t)),& \text{otherwise}
    \end{cases}
    \\
\end{equation}
for index $k$ over the set of positive integers and where each higher harmonic introduces four additional moduli. Subtracting the state B force response and normalizing by length scale $\Delta L$ and force scale $F_{A0}-F_{B0}$ result in two difference of moduli of the $k^{\rm th}$ harmonic: $\Delta e_k'  = \Delta L (E_{A,k}'-E_{B,k}')/(F_{A0}-F_{B0})$ and $\Delta e_k'' = \Delta L (E_{A,k}''-E_{B,k}'')/(F_{A0}-F_{B0})$. Expanding equations~\eqref{eqn:nondimensional}--\eqref{eqn:Wideal} to include the higher harmonics results in
\begin{align}
f(\phi)  &= 
    \begin{cases}
        1   + \sum\limits_k \left( \frac{w'_k }{w_0} \frac{l_B-l_A}{\alpha_k}\sin k\phi 
        + \frac{ w''_k }{w_0}\frac{l_B-l_A}{\beta_k} \cos k\phi \right), &  \text{for } \phi \in [\phiA, \phiD]\\
        0, & \text{otherwise}.
    \end{cases}
\\
w'_{k} &= -\Delta e'_k \int\limits_{\phiA}^{\phiD} \sin k\phi\cos\phi\, d\phi = -\Delta e'_k \alpha_k
\label{eqn:Wsprings,k}
\\
w''_{k} &= -\Delta e_k'' \int\limits_{\phiA}^{\phiD} \cos k\phi\cos\phi\, d\phi = -\Delta e_k'' \beta_k
\\
w_0  &=  -\int_{\ell_A}^{\ell_D} d\ell = \ell_A - \ell_B,
\end{align}
where $\alpha_{k} = \int_{\phiA}^{\phiD}\sin k\phi\cos\phi\, d\phi$ and  $\beta_{k} = \int_{\phiA}^{\phiD}\cos k\phi\cos\phi\, d\phi$ are shape parameters. The terms $w'_k$ and $w''_k$ are the work by storage forces and loss forces of the  $k^{\rm th}$ harmonic, respectively. The net work now includes contributes from all higher harmonics as 
\begin{equation}
    w_n = w_0 + \textstyle\sum\limits_k (w'_k + w''_k).
\end{equation}
The relation between nondimensional net work $w_n$ and the dimensional net work remains unchanged according to equation~\eqref{eqn:dimensional net work} because only the first harmonic of the state B force response contributes to net work.

\subsection{Traversing the shape space}
\label{Methods:Maxwell}
We document here the process of generating dimensionless spliced loops illustrated in \fig~\ref{fig:LoopSpace}c for a tunable Maxwell body in parallel with an ideal force generator. The force response to an oscillatory motion of unit amplitude is given by \eqn s~(\ref{eqn:nondimensional}-\ref{eqn:Wideal}) with $\phiA = \pi/6$ and $\phiD = 7\pi/6$. The $\Delta e'$ and $\Delta e''$ needed to calculate $w'$ and $w''$ are derived for a Maxwell body, a spring with unit stiffness in series with a dashpot with unit damping coefficient. Specifically, $\Delta e' = \omega^2 / (\omega^2 + 1)$ and $\Delta e'' = \omega   / (\omega^2 + 1)$.
The oscillatory frequencies $\omega$ are chosen such the spliced loops are uniformly spread on the semicircle on the Nyquist plot. Lastly, we hold the ideal motor work constant at $w_0 = 1.5$.

\subsection{Sculpin Work Loops}
\label{Methods:Sculpin}

We estimated $F_{A0}$, $E'_A$, and $E''_A$ in \eqn~\eqref{eqn:force interpolation fish} from the set of 12 work loops of a short-horned sculpin abdominal muscle published by Johnson and Johnston \citep{Johnson1991aa}.
All optimization and root-finding are performed using the Python \texttt{scipy.optimize} library \citep{2020SciPy-NMeth}.
The data were extracted by digitizing the original figures.

The interpolation variable $a(t)$ is found using published isometric twitch response of the same muscle, but separately measured by Altringham and Johnston \citep{Altringham1988aa} at a temperature between $2.5^\circ$C and $3.5^\circ$C. The work loop data were measured at $4^\circ$C.
A smooth approximation $ct^k\exp(-t/\tau_a)$ is fitted to the twitch response, which captures the fast initial rise and slower decay in the measurements.
When the twitch response is scaled to be between 0 and 1 where 1 is equal to the peak tension measured, we find these values to be $c = 3.39\cdot 10^{-4}$, $k = 2.99$, and $\tau_a = 13.2$ms.
To simulate the response to three electrical stimuli instead of one, we perform the following procedure:
{\bf 1.} Superimpose the three twitch responses separated by 20ms intervals.
{\bf 2.} Scale the superimposed response by its maximum so that interpolation is normalized to be between 0 and 1. We refer to this intermediate step as $a^*(t)$.
{\bf 3.} Impose periodicity by finding a baseline shift $\delta$ such that $a^*(\delta)$ = $a^*(T + \delta)$ where $T = 200$ms is the period of the sculpin work loops and define the desired output $a(t)$ to be $a^*(t + \delta)$.
The third step makes a minor correction to ensure that $a(t)$ is periodic and is necessary because the smooth fit to the data $ct^k\exp(-t/\tau_a)$ is not automatically periodic.
This is because, under steady periodic conditions, the value at $t=0$ is not zero.
So we use the initial time offset $\delta = 3.21$ms, which captures a non-zero initial condition of $a(0) = a(T) = 0.00356$.
The final result $a(t)$ is plotted in as an inset in \fig~\ref{fig:FishWorkLoops}c.

Given  $a(t)$, measured force response $y_i(t)$, and stimulus timing $\theta_i$ of the $i^{\rm th}$ sculpin work loop, a set of values ($F_{A0},E'_A, E''_A$) were generated for the $i^{\rm th}$ loop by excluding it and minimizing the error for the remaining loops.
So the fitted values for the $i^{\rm th}$ loop are obtained by minimizing
\begin{equation}
\sum_{m\neq i}\sum_{k}  \left[ a(t_k-\theta_m)(F_{A0} + \Delta L E'_A \sin(\omega t_k) + \Delta L E''_A \cos(\omega t_k)) - y_m(t_k) \right]^2
\end{equation}
where $k$ indexes the time steps in $[0,T]$.

\subsubsection{Goodness of fit measures}%
\label{supp:sec:Goodness of fit}
Pearson's correlation coefficient for the $i^{\rm th}$ loop is calculated from the fitted values as 
\begin{equation}
r = \dfrac{\sum_k (y_i(t_k)-\bar{y_i}) (F(t_k)-\bar{F})}{\sum_k (y_i(t_k)-\bar{y_i})\sum_k (F(t_k)-\bar{F})}
\end{equation}
where bars denote mean values and $F(t_k)$ is predicted force response in \eqn~\eqref{eqn:force interpolation fish} given the fitted values ($F_{A0},E'_A, E''_A$).
We use Pearson's correlation coefficient because of the well-known problems in assessing goodness of fit using the coefficient of determination $R^2$ that is employed for linear statistical models \citep{Kvalseth1985aa}.

\subsection{Sarcomere model simulation}
\label{Methods:MonteCarlo}
We use a published crossbridge model to assess the applicability of splicing \citep{Walcott2014aa}.
The attachment rate $f$ of a crossbridge to an internal displacement in the interval $[x,x + dx]$ is given by
\begin{equation}
    f(x,i,j,\varepsilon) = \beta(i,j,\varepsilon) \sqrt{\frac{D}{2\pi}}\exp\left(-\frac{D}{2}x^2\right)
\end{equation}
where $D = 100$. The function $\beta(i,j,\varepsilon)$ depends on coupling strength $\varepsilon$ and the distance to the nearest attached crossbridge on the left $i$ and right $j$ as
\begin{equation}
    \beta(i,j,\varepsilon) =
    \begin{cases}
        1 &\text{ if}\quad i+j \leq C\\
        \varepsilon^{ (i+j-C)/C} &\text{ if}\quad i,j<C \& i+j > C\\
        \varepsilon^{ i/C} &\text{ if}\quad i < C \& j \geq C\\
        \varepsilon^{ j/C} &\text{ if}\quad i \geq C \& j < C\\
        \varepsilon &\text{ if}\quad i \geq C \& j \geq C
    \end{cases}
\end{equation}
where $C =11$. The detachment rate $g$ of a crossbridge with internal displacement $x$ is 
\begin{equation}
    g(x) = \frac{1}{K}\exp(-E(x+1))
\end{equation}
where $E = 1$ and $K = 0.2$. 

A Monte Carlo simulation computed the behavior of 500,000 crossbridges as the system is sheared by an oscillatory motion $L(t) = A\sin(\omega t)$ with $A = 1$ and a coupling strength $\varepsilon$ that periodically alternates between $0.15$ and $\exp(-5)$, which respectively corresponds to pCa values of 6.17 and 7.04 \citep[via Walcott's \eqn~23]{Walcott2014aa}.
At each timestep, crossbridge attachment and detachment are decided by a random number generator. The tension $T_m$ per crossbridge transmitted across the system at the $m^{th}$ timestep is computed as 
\begin{equation}
    T_m = \frac{1}{N}\sum_{n = 1}^{N} (x_n + 1)s
\end{equation}
for index $n$ over $N$ crossbridges and where $s = 1$ if a crossbridge is attached and zero if detached.

All equations are nondimensional. The chosen timescale is $1/k_a$ where $k_a = 40$Hz is the attachment rate of a crossbridge with many attached neighbors. The length scale is a powerstroke distance $d = 10$nm. The stiffness scale is $k = 0.4$pN/nm, the stiffness of each crossbridge's elastic spring. Consequently, the tension scale is $kd = 4$pN in units of force per crossbridge. The Monte Carlo simulations were performed using Matlab version 9.8.0.1323502 (R2020a, Natick, MA). 

\subsection{Illustrative phenomenological muscle model}
\label{Appendix:nonlinear muscle model}
We use a previously published phenomenological muscle model \citep{Todorov2002aa} to illustrate the splicing between nonlinear rheologies shown in \fig~\ref{fig:Geometric}a. The force output $F(a,l,v)$ is a function of activation $a$, length $l$ and velocity $v$ according to,
\begin{equation}
    F(a,l,v) = B(a,l) F_L(l)F_V(l,v) + F_{P1}(l) + BF_{P2}(l),
\end{equation}
where the functions are defined as,
    \begin{align}
        B(a,l)   &= 1-\exp\left(\left(-\frac{a}{0.56N_f(l)}\right)^{N_f(l)}\right),  \\  
        N_f(l)   &= 2.11 + 4.16(l^{-1}-1),\\
        F_L(l)   &=\exp\left(-\left|\frac{l^{1.93} -1}{1.03}\right|^{1.87}\right) ,  \\
        F_V(l,v) &= 
        \begin{cases}
            \frac{-5.72-v}{-5.72 + (1.38 + 2.09l)v}, \quad v\leq 0 \\
            \frac{0.62 - (-3.12 + 4.21l  -2.67l^2)v}{0.62 + v}, \quad v >  0
        \end{cases} \\
        F_{P1}(l)&= 5.42\ln\left(\exp\left(\frac{l-1.42}{0.052}\right)+1\right),\\
        F_{P2}(l)&= -0.02\exp(-18.7(l-0.79)-1).
    \end{align}

To generate the nonlinear force responses, $a=1$ is used as the first activation state and $a=0.5$ as the second. The dimensional length input is $l(t) = 1 + 0.15\sin(( 3\pi/2) t)$, activation phase is $\phi_A = \sin^{-1}(0.5)$, and deactivation phase is $\phi_D = (\sin^{-1}(0.4)+\pi)$. All simulations were performed using Matlab version 9.8.0.1323502 (R2020a, Natick, MA). 

\section*{List of source files}
{\bf Source code}: Matlab code used to generate the sarcomere model described in Methods \ref{Methods:MonteCarlo} and used in \fig~\ref{fig:MonteCarlo}.
\newpage

\renewcommand{\thefigure}{\ref{fig:Ellipse} - Figure Supplement 1}
\begin{figure}[p]
    \centering
    \includegraphics[width=\linewidth]{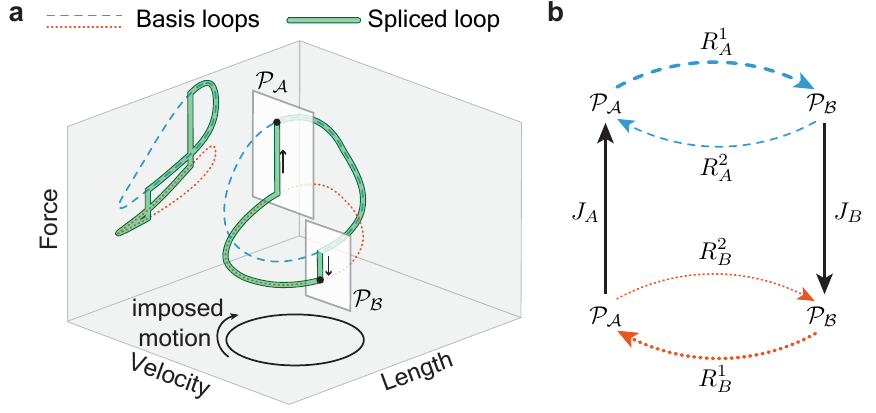}
    \caption{
	The spliced loop can be viewed as a singular periodic orbit of a piecewise smooth dynamical system with two switching planes  $\mathcal{P_B}$ and $\mathcal{P_A}$ that are defined by the phases $\phiD$ and $\phiA$ when the rheological state is changed from $A$ to $B$ or {\it vice versa}.
	At rheological states $B$ or $A$, the dynamics of the material when subjected to periodic length oscillations are governed by the composite functions $(R_B^1 \circ R_B^2)$ and $(R_A^2 \circ R_A^1)$, respectively, which map initial conditions on the plane $\mathcal{P}_A$ to the plane $\mathcal{P}_B$ and back onto $\mathcal{P}_A$ again.
	The oscillatory responses at states, $B$ or $A$, correspond to stable periodic orbits of the two dynamical systems.
    The existence of the orbit implies that a steady response can be measured under a constant rheological state.
	The splicing of those two periodic orbits by instantaneously switching between them at the planes $\mathcal{P_B}$ and $\mathcal{P_A}$ yields a new periodic orbit for the spliced dynamical system $(J_D \circ R_A^1 \circ J_A \circ R_B^1)$.
	The instantaneous jumps $J_A$ and $J_B$ between slowly varying trajectories resembles approaches from geometric singular perturbation theory for constructing relaxation oscillations in multiple-timescale systems \citep{Bernardo2008aa,Grasman2011aa}.
    A widely used phenomenological Hill-type muscle model \citep[Methods \ref{Appendix:nonlinear muscle model},][]{Todorov2002aa} was used to illustrate the nonlinear, and non-elliptical basis loops.
    }
    \label{fig:Geometric}
\end{figure}

\renewcommand{\thetable}{Figure \ref{fig:FishWorkLoops} - Table Supplement 1}
\begin{table}[p]
    \centering
    \caption{Fitted ellipses to the sculpin work loop data \citep{Johnson1991aa}. The scale $F_{\rm max}$ is the difference between the maximal and minimal force in the $5^\circ$ work loop}
    \label{tab:fitting ellipses to work loops}
    \includegraphics[width=\textwidth]{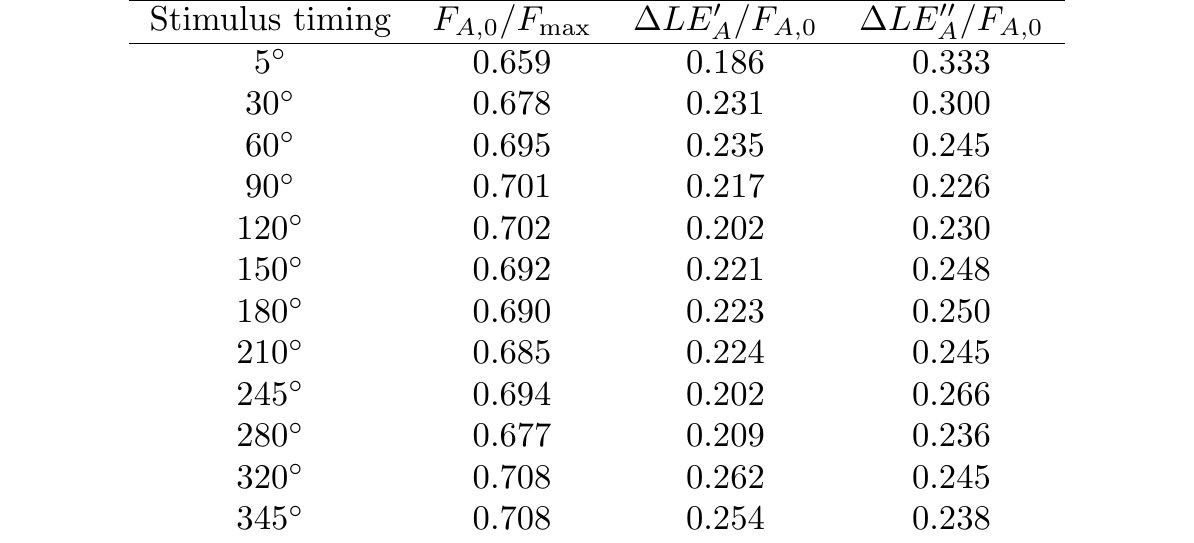}
\end{table}%

\renewcommand{\thefigure}{\ref{fig:FishWorkLoops} - Figure Supplement 1}
\begin{figure}[p]
    \centering
    \includegraphics[width=\linewidth]{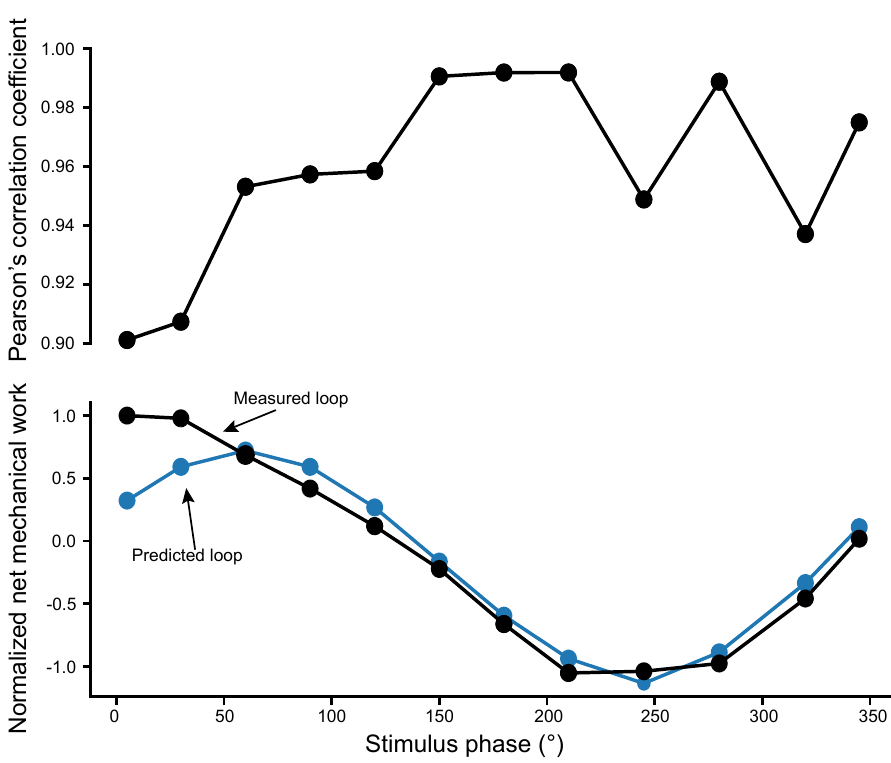}
    \caption{
    Quantitative comparisons between measured sculpin work loops and loops predicted from splicing hypothesis.
    (Top) The Pearson's correlation is calculated for each pair of measured and predicted loop as a function of the stimuli timing in units of degrees from 0$^\circ$ to 360$^\circ$.
    See Methods \ref{Methods:Sculpin} for details of the calculation.
    (Bottom) The net mechanical work is calculated for both the measured and predict loops by taking the total enclosed area of each loop. The values are normalized by the net work performed by the measured 5$^\circ$ loop.
    }
    \label{fig:FishWorkLoopComparison}
\end{figure}

\renewcommand{\thefigure}{\ref{fig:MonteCarlo} - Figure Supplement 1}
\begin{figure}[p]
    \centering
    \includegraphics[width=\linewidth]{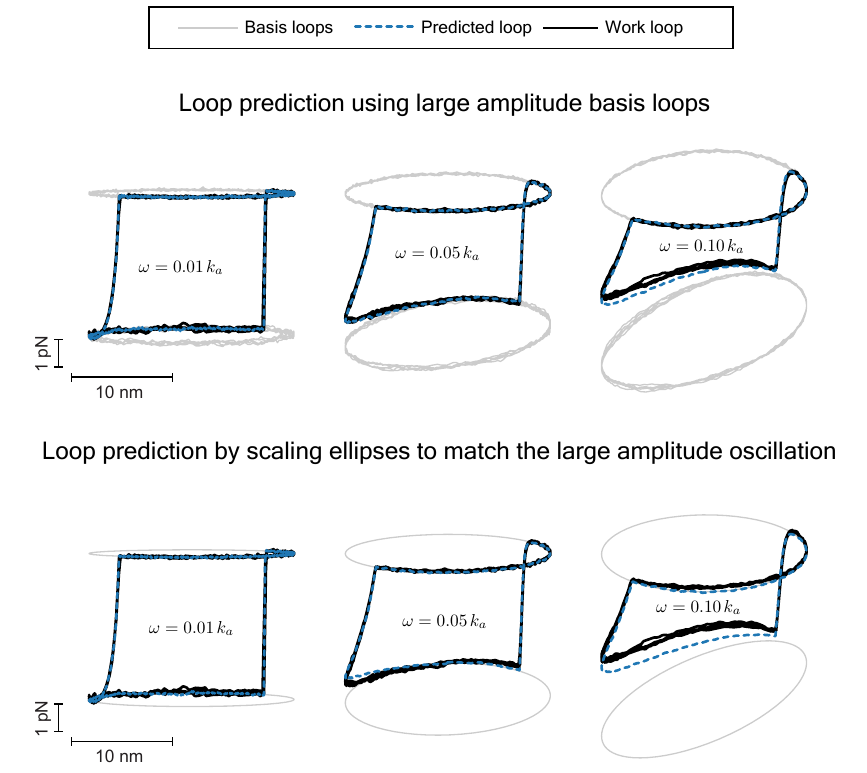}
    \caption{ Illustration of  how small-amplitude basis loops can be used to generate large-amplitude work loop predictions.
    When large amplitude basis loops are inaccessible to direct measurement, scaling up linear ellipses found from small amplitude oscillations results in minimal loss of accuracy between predicted loops and work loops for the sarcomere model. The work loops (solid black lines) are the same in both panels and serve as a reference for comparison of predicted loops. 
    All loops are generated from Monte Carlo simulations of 500,000 crossbridges following the flow chart of the main figure.
    }
    \label{fig:SmallAmplitude}
\end{figure}

\end{document}